\documentclass[runningheads]{llncs}
\usepackage{amsfonts}
\usepackage{graphicx}
\usepackage{amsfonts}
\usepackage{graphicx}
\usepackage{epsfig}
\usepackage{graphicx}
\usepackage{amsmath}
\usepackage{amssymb}
\usepackage{mathtools}
\usepackage{resizegather}
\usepackage{marvosym}
\usepackage{dsfont}
\usepackage{enumitem}
\usepackage{caption}
\usepackage{multirow}
\RequirePackage[utf8]{inputenc}

\usepackage[colorlinks=true, allcolors=blue]{hyperref}

\usepackage[ruled, vlined]{algorithm2e}

\usepackage{lipsum}
\usepackage{array}
\usepackage{amsmath}

\usepackage{enumitem}
 
\begin{document}

\title{Speech Audio Synthesis from Tagged MRI and Non-Negative Matrix Factorization via Plastic Transformer} 

\author{Xiaofeng Liu\inst{1} \and Fangxu Xing\inst{1}  \and Maureen Stone\inst{2} \and Jiachen Zhuo\inst{2} \and Sidney Fels\inst{3} \and Jerry L. Prince\inst{4} \and  Georges El Fakhri\inst{1} \and Jonghye Woo\inst{1}}

\titlerunning{Plastic Transformer for 4D Tagged-MRI2Audio via NMF}

\institute{Massachusetts General Hospital and Harvard Medical School, Boston, MA, USA\and
University of Maryland, Baltimore, MD, USA \and
University of British Columbia, Vancouver, BC, Canada \and
Johns Hopkins University, Baltimore, MD, USA
}
 
\authorrunning{X. Liu et al.}

\maketitle              % typeset the header of the contribution

\begin{abstract}
The tongue's intricate 3D structure, comprising localized functional units, plays a crucial role in the production of speech. When measured using tagged MRI, these functional units exhibit cohesive displacements and derived quantities that facilitate the complex process of speech production. Non-negative matrix factorization-based approaches have been shown to estimate the functional units through motion features, yielding a set of building blocks and a corresponding weighting map. Investigating the link between weighting maps and speech acoustics can offer significant insights into the intricate process of speech production. To this end, in this work, we utilize two-dimensional spectrograms as a proxy representation, and develop an end-to-end deep learning framework for translating weighting maps to their corresponding audio waveforms. Our proposed plastic light transformer (PLT) framework is based on directional product relative position bias and single-level spatial pyramid pooling, thus enabling flexible processing of weighting maps with variable size to fixed-size spectrograms, without input information loss or dimension expansion. Additionally, our PLT framework efficiently models the global correlation of wide matrix input. To improve the realism of our generated spectrograms with relatively limited training samples, we apply pair-wise utterance consistency with Maximum Mean Discrepancy constraint and adversarial training. Experimental results on a dataset of 29 subjects speaking two utterances demonstrated that our framework is able to synthesize speech audio waveforms from weighting maps, outperforming conventional convolution and transformer models. 

%Yet, directly mapping two-dimensional weighting maps to their corresponding one-dimensional waveforms is challenging.

 %Because of their heterogeneous representations, however, direct mapping between the two modalities---i.e., two-dimensional (mid-sagittal slice) plus time tagged-MRI sequence and its corresponding one-dimensional waveform---is not straightforward. Instead, we resort to two-dimensional spectrograms as an intermediate representation, which contains both pitch and resonance, from which to develop an end-to-end deep learning framework to translate from a sequence of tagged-MRI to its corresponding audio waveform with limited dataset size.~Our framework is based on a novel fully convolutional asymmetry translator with guidance of a self residual attention strategy to specifically exploit the moving muscular structures during speech.~In addition, we leverage a pairwise correlation of the samples with the same utterances with a latent space representation disentanglement strategy.~Furthermore, we incorporate an adversarial training approach with generative adversarial networks to offer improved realism on our generated spectrograms.~Our experimental results, carried out with a total of 63 tagged-MRI sequences alongside speech acoustics, showed that our framework enabled the generation of clear audio waveforms from a sequence of tagged-MRI, surpassing competing methods. Thus, our framework provides the great potential to help better understand the relationship between the two modalities.

 \end{abstract}

\section{Introduction}

%The multimodal data,  that describe the same event can provide complementary information to under

Intelligible speech is produced by the intricate three-dimensional structure of the tongue, composed of localized functional units~\cite{woo2021deep}. These functional units, when measured using tagged magnetic resonance imaging (MRI), exhibit cohesive displacements and derived quantities that serve as intermediate structures linking tongue muscle activity to tongue surface motion, which in turn facilitates the production of speech. A framework based on sparse non-negative matrix factorization (NMF) with manifold regularization can be used to estimate the functional units given input motion features, which yields a set of building blocks (or basis vectors) and a corresponding sparse weighting map (or encoding)~\cite{woo2020identifying}. The building blocks can form and dissolve with remarkable speed and agility, yielding highly coordinated patterns that vary depending on the specific speech task at hand. The corresponding weighting map can then be used to identify the cohesive regions and reveal the underlying functional units~\cite{woo2018sparse}. As such, by elucidating the relationship between the weighting map and intelligible speech, we can gain valuable insights for the development of speech motor control theories and the treatment of speech-related disorders.

%However, heterogeneous data representations between two-dimensional (2D) weighting maps and high-frequency one-dimensional (1D) audio waveform make their translation a challenging task \cite{chung2016lip,akbari2018lip2audspec}. Besides, cross-modality speech models often lose pitch information \cite{akbari2018lip2audspec,ephrat2017vid2speech}. In contrast, a 2D spectrogram converted from its 1D audio waveform represents the energy distribution of audio signals over the frequency domain along the time axis, which contains both pitch and resonance information of audio signals \cite{akbari2018lip2audspec,ephrat2017vid2speech,liu2022cmri2spec}. Then, mel spectrograms can be easily converted back into audio waveforms \cite{griffin1984signal}.

%%%%%%%%%%%%%%%%%%%%%%%%%%%%%
%To facilitate our understanding of speech motor control in healthy and disease populations, associating dynamic imaging data with speech audio waveforms is an essential step in identifying the underlying relationship between tongue and oropharyngeal muscle deformation and its corresponding acoustic information \cite{liu2022cmri2spec}. Naturally, audio data recorded during scanning sessions need to be strictly paired and temporally synced with the dynamic tagged-MRI data to maintain their underlying relationship. In addition, the missing, noise-corrupted, or un-matched audio of the imaging data is common in practice. Therefore, reconstructing audio from imaging data itself becomes a necessary topic to be explored.

Despite recent advances in cross-modal speech processing, translating between varied-size of wide 2D weighting maps and high-frequency 1D audio waveforms remains a challenge. The first obstacle is the inherent heterogeneity of their respective data representations, compounded by the tendency of losing pitch information in audio \cite{chung2016lip,akbari2018lip2audspec}. By contrast, transforming a 1D audio waveform into a 2D spectrogram provides a rich representation of the audio signal's energy distribution over the frequency domain, capturing both pitch and resonance information along the time axis \cite{ephrat2017vid2speech,he2020image2audio}. Second, the input sizes of the weighting maps vary between 20$\times$5,745 and 20$\times$11,938, while the output spectrogram has a fixed size for each audio section. Notably, fully connected layers used in \cite{akbari2018lip2audspec} require fixed size input, while the possible fully convolution neural networks (CNN) can have varied output sizes and unstable performance~\cite{richter2021input}. Third, modeling global correlations for the long column dimension of the weighting map and the lack of spatial local neighboring relationships in the row dimension presents further difficulties for conventional CNNs that rely on deep hierarchy structure for expanding the reception field~\cite{luo2016understanding,araujo2019computing}. Furthermore, the limited number of training pairs available hinders the large model learning process.

To address the aforementioned challenges, in this work, we propose an end-to-end translator that generates 2D spectrograms from 2D weighting maps via a heterogeneous plastic light transformer (PLT) encoder and a 2D CNN decoder. The lightweight backbone of PLT can efficiently capture the global dependencies with a wide matrix input in every layer~\cite{huang2022lightvit}. Our PLT module is designed with directional product relative position bias and single-level spatial pyramid pooling to enable flexible global modeling of weighting maps with variable sizes, producing fixed-size spectrograms without information loss or dimension expansion due to cropping, padding, or interpolation for size normalization. To deal with a limited number of training samples, we explore pair-wise utterance consistency as prior knowledge with Maximum Mean Discrepancy (MMD)~\cite{dziugaite2015training} in a disentangled latent space as an additional optimization objective. Additionally, a generative adversarial network (GAN) \cite{goodfellow2020generative} can be incorporated to enhance the realism of the generated spectrograms.

The main contributions of this work are three-fold:

%To our knowledge, this is the first attempt at relating functional units via weighting maps derived from tagged MRI and sparse NMF with manifold regularization with audio waveforms via spectrograms.

\noindent$\bullet$ To our knowledge, this is the first attempt at relating functional units with audio waveforms by means of intermediate representations, including weighting maps and spectrograms.

\noindent$\bullet$ We developed a plastic light-transformer to achieve efficient global modeling of position sensitive weighting maps with variable sizes and long dimensions. 

\noindent$\bullet$ We further explored the pair-wise utterance consistency constraint with MMD minimization and adversarial training as additional supervision signals to deal with relatively limited training samples.

Both quantitative and qualitative evaluation results demonstrate superior synthesis performance over comparison methods. Our framework has the potential to support clinicians and researchers in deepening their understanding of the interplay between tongue movements and speech waveforms, thereby improving treatment strategies for patients with speech-related disorders.

\section{Methods}

\begin{figure}[t]
\begin{center}
\includegraphics[width=1\linewidth]{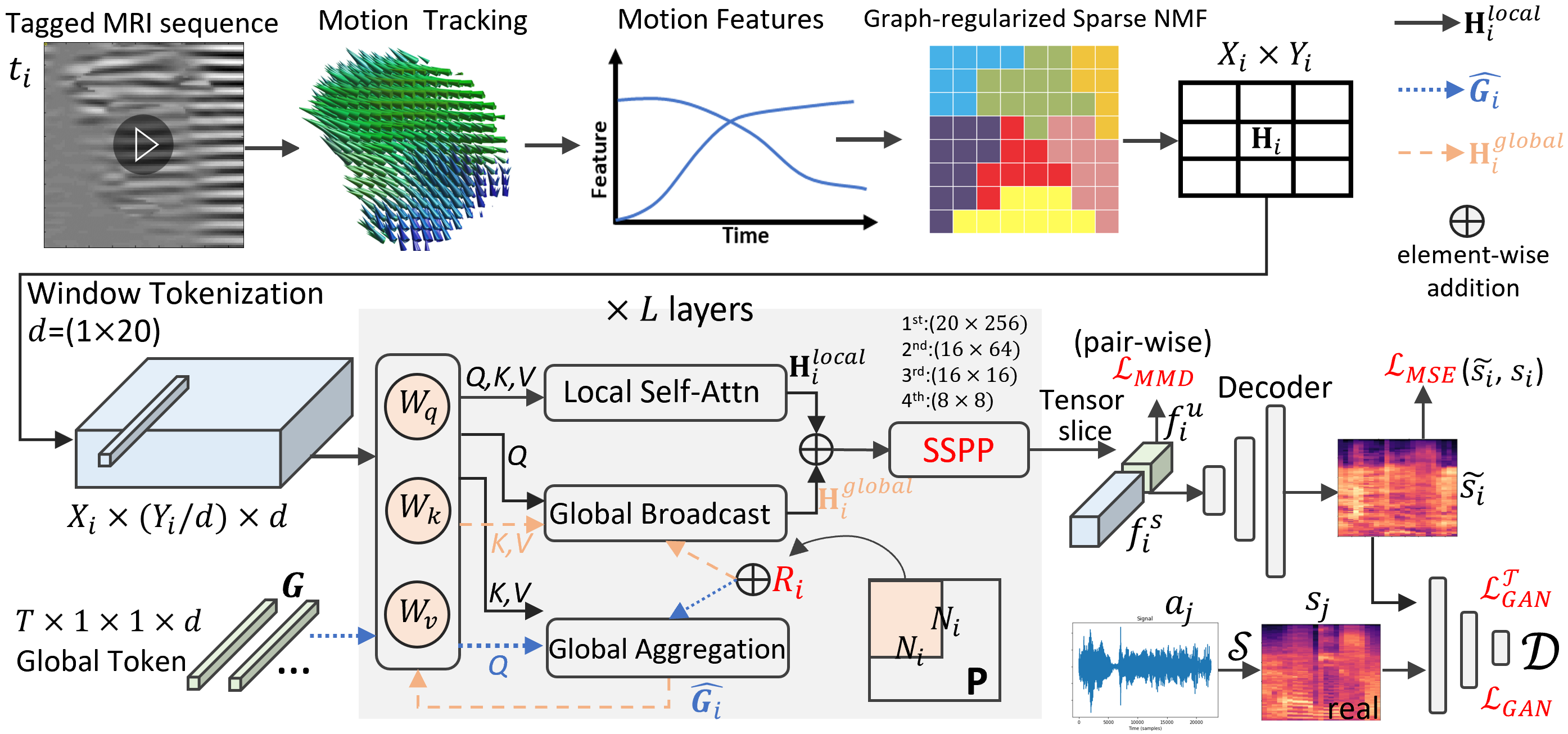}    
\end{center} 
\caption{Illustration of our translation framework. Only the NMF and translator with heterogeneous PLT encoder and 2D CNN decoder are used for testing.} 
\label{fig:illus} 
\end{figure} 

\subsection{Preprocessing}
During the training phase, we are given $M$ pairs of synchronized tagged MRI sequences $t_i$ and audio waveforms $a_i$, i.e., $\{t_i,a_i\}_{i=1}^M$. First, we apply a non-linear transformation using \href{https://librosa.org/doc/main/generated/librosa.feature.melspectrogram.html.}{librosa} to convert $a_i$ into mel-spectrograms, denoted as $s_i$ with the function $\mathcal{S}:a_i\rightarrow s_i$. This transformation uses a Hz-scale to emphasize human voice frequencies ranging from 40 to 1000 Hz, while suppressing high-frequency instrument noise. Second, for each tagged MRI sequence $t_i$, we use a phase-based diffeomorphic registration method~\cite{xing2017phase} to track the internal motion of the tongue. This allows us to generate corresponding weighting maps denoted as ${\bf{H}}_i$, which are based on input motion features $\mathbf{X}_i$, including the magnitude and angle of each track, by optimizing the following equation.
\begin{equation}
\mathcal{E} = \frac{1}{2}{\left\| {\mathbf{X}_i - \mathbf{W}_i\mathbf{H}_i} \right\|_F^2 + \frac{1}{2} \lambda \mathrm{Tr}(\mathbf{H}_i \mathbf{L}_i \mathbf{H}_i^\top) + \eta \left\| \mathbf{H}_i \right\|_{1/2} }, % \right],
\label{eq:eq10}
\end{equation}
where $\lambda$ and $\eta$ denote the weights associated with the manifold and sparse regularizations, respectively, and Tr($\cdot$) represents the trace of a matrix. The graph Laplacian is denoted by $\mathbf{L}$.

\subsection{Encoding variable size $\bf{H}_i$ with plastic light-transformer}
 
Directly modeling correlations among any two elements in a given weighting map ${{\bf{H}}_i}\in\mathbb{R}^{X_i\times Y_i}$ can impose quadratic complexity of $\mathcal{O}(X_i^2Y_i^2)$.
The recent efficient vision transformers (ViTs)~\cite{huang2022lightvit,liu2021swin,zhang2021rest,liu2021swin,chu2021twins} usually adopt a local patch design to compute local self-attention and correlate patches with CNNs. Specifically, the input is divided into $N_i=\frac{X_i}{P_x}\times\frac{Y_i}{P_y}$ patches\footnote{The bottom-right boundary is padded with 0 to ensure $X_i\%{P_x}=0$ and $Y_i\%{P_y}=0$.}, each of which is flattened to a token vector with a length of $d=P_x\times P_y$~\cite{dosovitskiy2020image}. The local self-attention is then formulated with a complexity of $\mathcal{O}(N_id^2=X_iY_id)$ as follows: 
\begin{align}
   {{\bf{{H}}}_i^{\mathrm{local}}}=\mathrm{Attn}({\bf{{H}}}_i^q,{\bf{{H}}}_i^k,{\bf{{H}}}_i^v)=\mathrm{SoftMax}(\frac{{{\bf{{H}}}_i^q}{{\bf{{H}}}_i^{k\top}}}{\sqrt{d}}){{\bf{{H}}}_i^v}, \in\mathbb{R}^{X_i\times Y_i},
\end{align}where vectors ${\bf{{H}}}_i^q$, ${\bf{{H}}}_i^k$, ${\bf{{H}}}_i^v\in\mathbb{R}^{N_i\times d}$ are produced by the linear projections of query ($W_q$), key ($W_k$), and value ($W_v$) branches, respectively~\cite{dosovitskiy2020image,zhang2021rest,chu2021twins}. The global correlation of ViTs with CNN~\cite{zhang2021rest,chu2021twins,yang2021focal} or window shifting~\cite{liu2021swin}, however, may not be efficient for our wide matrix ${{\bf{H}}_i}$, which lacks explicit row-wise neighboring features and may have a width that is too long for hierarchical convolution modeling. To address these challenges, we follow the lightweight ViT design~\cite{huang2022lightvit}, which uses a global embedding ${\bf{G}}\in\mathbb{R}^{T\times d}$ with $T\ll N_i$ randomly generated global tokens as the anchor for global information aggregation ${\bf{\hat{G}}}_i$. The aggregation is performed with attention of ${{\bf{{G}}}^q,{\bf{H}}_i^k,{\bf{H}}_i^v}$, which is then broadcasted with attention of ${{\bf{{H}}}_i^q,{\bf{\hat{G}}}_i^k,{\bf{\hat{G}}}_i^v}$ to leverage global contextual information~\cite{huang2022lightvit}. 

 %Therefore, we follow light ViT design~\cite{huang2022lightvit} to utilize a global embedding ${\bf{G}}\in\mathbb{R}^{T\times d}$ with $T$ random generated $d$-dim global tokens as the anchor for global information aggregation ${\bf{\hat{G}}}_i$ with attention of $\{{\bf{{G}}}^q,{\bf{H}}_i^k,{\bf{H}}_i^v\}$ and then broadcast with attention of $\{{\bf{{H}}}_i^q,{\bf{\hat{G}}}_i^k,{\bf{\hat{G}}}_i^v\}$~\cite{huang2022lightvit}. 

While LightViT backbones have been shown to achieve wide global modeling within each layer~\cite{huang2022lightvit}, they are not well-suited for our variable size input and fixed size output translation. Although the self-attention scheme used in ViTs does not constrain the number of tokens, the absolute patch-position encoding in conventional ViTs~\cite{dosovitskiy2020image} can only be applied to a fixed $N_i$~\cite{zhang2021rest}, and the attention module will keep the same size of input and output. Notably, the number of tokens $N_i$ will change depending on the size of $X_i\times Y_i$. As such, in this work, we resort to the directional product relative position bias~\cite{wu2021rethinking} to add ${\bf{R}}_i\in\mathbb{R}^{N_i\times N_i}$, where element $r_{a,b}={\bf{p}}_{\delta^x_{a,b},\delta^y_{a,b}}$ is a trainable scalar, indicating the relative position weight between the patches $a$ and $b$\footnote{a learnable matrix ${\bf{p}}\in\mathbb{R}^{(2P_x-1)\times(2P_y-1)}$ is initialized with trunc\_normal\_, where $P_x=20$ and $P_y=\frac{12000}{20}=600$ are the maximum patch dimensions in our task.}. We set the offset of patch position in $x$ and $y$ directions $\delta^x_{a,b}= x_a-x_b+P_x, \delta^y_{a,b}=y_a-y_b+M_y$ as the index in ${\bf{p}}$. Furthermore, the product relative position bias utilized in this work can distinguish between vertical or horizontal offsets, whereas the popular cross relative position bias~\cite{wu2021rethinking} in computer vision tasks does not need to differentiate between time and spatial neighboring relationships in two dimensions.

Therefore, for global attention, we can aggregate the information of local tokens by modeling their global dependencies with 
\begin{align}
   {\bf{\hat{G}}}_i=\mathrm{Attn}({\bf{{G}}}^q,{\bf{H}}_i^k,{\bf{H}}_i^v)=\mathrm{SoftMax}(\frac{{\bf{{G}}}^q{\bf{H}_i^{k\top}}+{\bf{R}}_i}{\sqrt{d}}){\bf{H}}_i^v, \in\mathbb{R}^{X_i\times Y_i}.
\end{align}
Then, these global dependencies are broadcasted to every local token: 
\begin{align}
   {{\bf{H}}_i^{\mathrm{global}}}=\mathrm{Attn}({\bf{{H}}}_i^q,{\bf{\hat{G}}}_i^k,{\bf{\hat{G}}}_i^v)=\mathrm{SoftMax}(\frac{{\bf{H}}_i^q {\bf{\hat{G}}_i^{k\top}}+{\bf{R}}_i}{\sqrt{d}}){\bf{\hat{G}}}_i^v, \in\mathbb{R}^{X_i\times Y_i}.
\end{align}
By adding ${{\bf{H}}_i^{\mathrm{local}}}$ and ${{\bf{H}}_i^{\mathrm{global}}}$, each token can benefit from both local and global features, while maintaining linear complexity with respect to the input size. This brings noticeable improvements with negligible FLOPs increment. However, the sequentially proportional patch merging used in \cite{huang2022lightvit,zhang2021rest,chu2021twins} still generates output sizes that vary with input sizes. Therefore, we utilize the single-level Spatial Pyramid Pooling (SSPP) \cite{he2015spatial} to extract a fixed-size feature for arbitrary input sizes. As illustrated in Fig. 1, the output of our channel-wise SSPP module with 20$\times$256 bins has the size of $20\times 256\times d$, which can be a token merging scheme that adapts to the input size. Therefore, the final output of a layer is given by 
\begin{align}
   {{\bf{H}}'_i}=\mathrm{SSPP}({{\bf{H}}_i^{\mathrm{local}}}+{{\bf{H}}_i^{\mathrm{global}}}) \in\mathbb{R}^{X_i\times Y_i}.
\end{align}
We cascade four PLT layers with SSPP as our encoder to extract the feature representation $f_i\in\mathbb{R}^{8\times8\times d}$. For the decoder, we adopt a simple 2D CNN with three deconvolutional layers to synthesize the spectrogram $\tilde{s}_i$.

\subsection{Overall Training Protocol}
 
We utilize the intermediate pairs of $\{{\bf{H}}_i,s_i\}_{i=1}^M$ to train our translator $\mathcal{T}$, which consists of a PLT encoder and a 2D CNN decoder. The quality of the generated spectrograms $\tilde{s}_i$ is evaluated using the mean square error (MSE) with respect to the ground truth spectrograms $s_i$: 
\begin{align}
    \mathcal{L}_{\mathrm{MSE}} = ||\tilde{s}_i-{s}_i||_2^2 = ||\mathcal{T}({\bf{H}}_i)-\mathcal{S}(a_i)||_2^2.
\end{align} 
%The intermediate pairs of $\{{\bf{H}}_i,s_i\}_{i=1}^M$ are utilized for the training of our translator $\mathcal{T}$ with a PLT encoder and a 2D CNN decoder. For the generated $\tilde{s}_i$, the mean square error (MSE) is applied to compare it with $s_i$:

Additionally, we utilize the utterance consistency in the latent feature space as an additional optimization constraint. Specifically, we propose to disentangle $f_i$ into two parts, i.e., utterance-related $f^u_i$ and subject-related $f^s_i$. In practice, we split the utterance/subject-related parts channel-wise using \href{https://pytorch.org/docs/stable/generated/torch.Tensor.select.html}{tensor slicing method}. Following the idea of deep metric learning \cite{liu2017adaptive}, we aim to minimize the discrepancy between the latent features $f^u_i$ and $f^u_j$ of two samples $t_i$ and $t_j$ that belong to the same utterance. Therefore, we use MMD~\cite{dziugaite2015training} as an efficient discrepancy loss $\mathcal{L}_{\mathrm{MMD}}=\gamma \mathrm{MMD}(f^u_i,f^u_j)$, where $\gamma=1$ or $0$ for same or different utterance pairs, respectively.

Of note, the $f^s_i$ is implicitly encouraged to incorporate the subject-related style of the articulation other than $f^u_i$ with a complementary constraint \cite{liu2019feature,liu2021mutual} for reconstruction. Therefore, the decoder, which takes $f^s_i$ conditioned on $f^u_i$ can be considered as the utterance-conditioned spectrogram distribution modeling. This approach follows a divide-and-conquer strategy \cite{che2019deep,liu2021domain} for each utterance and can be particularly efficient for relatively few utterance tasks.

A GAN model can be further utilized to boost the realism of $\tilde{s}_i$. A discriminator $\mathcal{D}$ is employed to differentiate whether the mel-spectrogram is real ${s}_i=\mathcal{S}(a_i)$ or generated $\tilde{s}_i=\mathcal{T}({\bf{H}}_i)$ with the following binary cross-entropy loss:  
\begin{align}
\mathcal{L}_{\mathrm{GAN}} = \mathbb{E}_{{s}_i}\{\text{log}(\mathcal{D}({s}_i))\} +  \mathbb{E}_{\tilde{s}_i}\{\text{log}(1-\mathcal{D}(\tilde{s}_i))\}.
\end{align}
In adversarial training, the translator $\mathcal{T}$ attempts to confuse $\mathcal{D}$ by optimizing $\mathcal{L}^{\mathcal{T}}_{GAN}= \mathbb{E}_{\tilde{s}_i}\{-\text{log}(1-\mathcal{D}(\tilde{s}_i))\}$. Of note, $\mathcal{T}$ does not involve real spectrograms in $\text{log}(\mathcal{D}(s'_i))$ \cite{salimans2016improved}. Therefore, the overall optimization objectives of our translator $\mathcal{T}$ and discriminator $\mathcal{D}$ are expressed as: 
\begin{align}
    ^{\text{min}}_{~\mathcal{T}}~ \mathcal{L}_{\mathrm{MSE}} + \beta\mathcal{L}_{\mathrm{MMD}}+ \lambda\mathcal{L}^{\mathcal{T}}_{\mathrm{GAN}};~~~^{\text{min}}_{~\mathcal{D}}~  \mathcal{L}_{\mathrm{GAN}},
\end{align}
where $\beta$ and $\lambda$ represent the weighting parameters. Notably, only $\mathcal{T}$ is utilized in testing, and we do not need pairwise inputs for utterance consistency. Recovering audio waveform from mel-spectrogram can be achieved by the well-established Griffin-Lim algorithm~\cite{griffin1984signal} in the \href{https://librosa.org/doc/main/generated/librosa.feature.inverse.mel_to_audio.html}{Librosa} toolbox.

\begin{figure*}[t]
\begin{center} 
\includegraphics[width=1\linewidth]{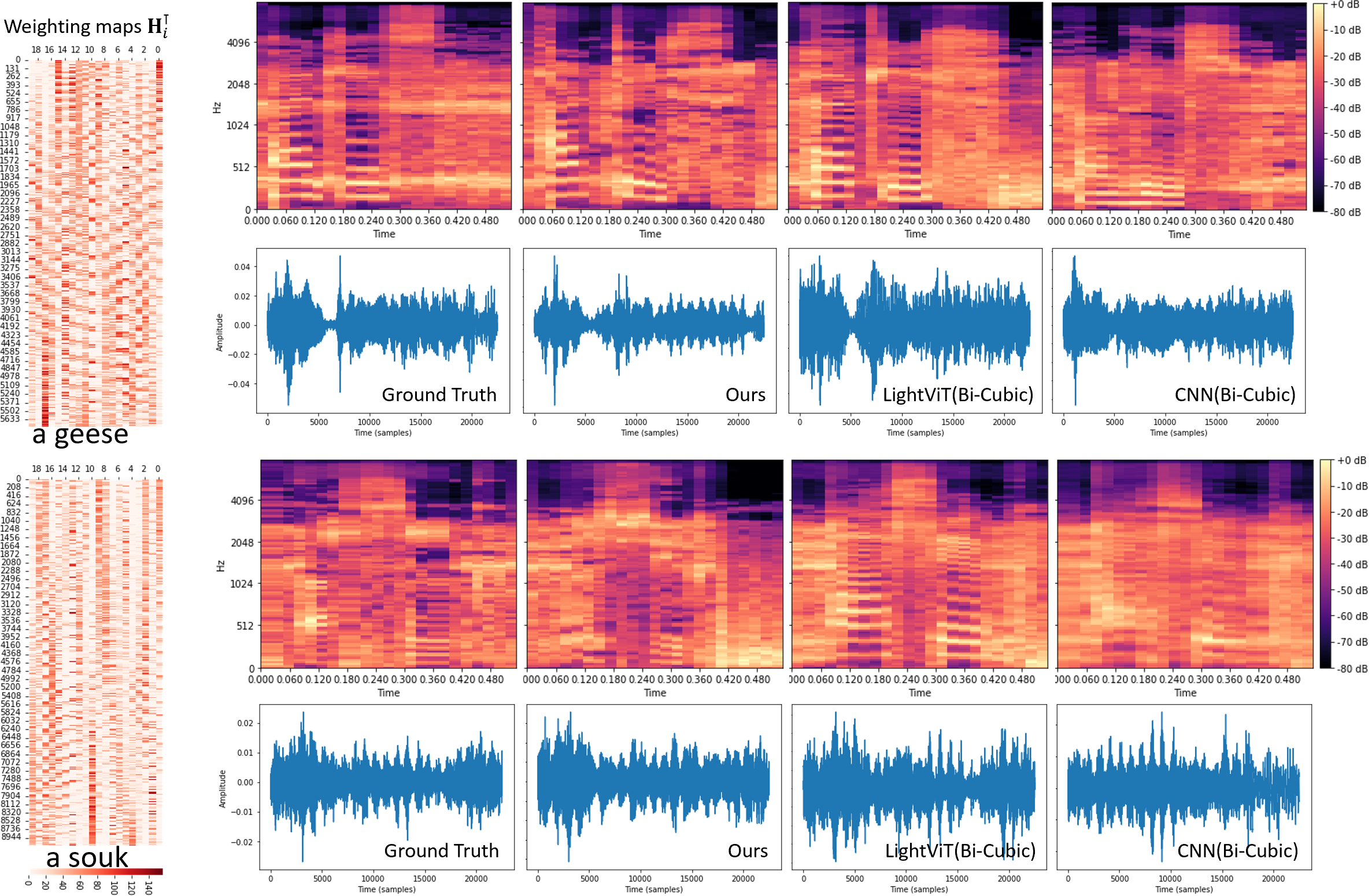}   
\end{center} 
\caption{Comparisons of our PLT with CNN and LightViT using bi-cubic interpolation. We show ${\bf{H}}_i^\top$ for compact layout. Audios are attached in supplementary.}  
\label{fig:results} 
\end{figure*}

\section{Experiments and Results}
 
For evaluation, we collected paired 3D tagged MRI sequences and audio waveforms from a total of 29 subjects, while performing the speech words ``a souk" or ``a geese," with a periodic metronome-like sound as guidance~\cite{lee2013semi,xing20133d}. The tagged-MRI sequences consisted of 26 frames, which were resized to 128$\times$128. The resulting $\bf{H}$ matrix varied in size from 20$\times$5,745 to 20$\times$11,938 (we set one dimension to a constant value of 20.) The audio waveforms had varying lengths between 21,832 to 24,175. To augment the dataset, we employed a sliding window technique on each audio, allowing us to crop sections with 21,000 time points, resulting in 100 audio waveforms. Then, we utilized the Librosa library to convert all audio waveforms into mel-spectrograms with a size of 64$\times$64. For our evaluation, we utilized a subject-independent leave-one-out approach. For the data augmentation of the $\bf{H}$ matrix, we randomly drop the column to round $Y_i$ to the nearest hundred, e.g., 9,882 to 9,800, generating 100 versions of $\bf{H}$. We utilized the leave-one-out evaluation, following a subject-independent manner. 

%Our data was acquired using a Siemens 3.0T TIM Trio system, with a 12-channel head coil and a 4-channel neck coil utilized for a segmented gradient echo sequence~\cite{lee2013semi,xing20133d}. The imaging parameters for the study were set as follows: the field of view was 240mm$\times$240mm on the mid-sagittal slice, with a resolution of 1.87mm$\times$1.87mm. 

%The best results are indicated in \textbf{bold}.
\begin{table}[t]
\centering 
\caption{Numerical comparisons during testing using leave-one-out evaluation} 
\resizebox{0.85\linewidth}{!}{
\begin{tabular}{l|c|c}
\hline
Encoder Models& Corr2D for spectrogram $\uparrow$ & PESQ for waveform $\uparrow$\\\hline\hline
 
%%???Lip2AudSpect~\cite{akbari2018lip2audspec} w/o NMF	& ?? & ??  \\\hline

CNN (Crop) 	&0.614$\pm$0.013  &  1.126$\pm$0.021 \\

CNN (Padding 0) &0.684$\pm$0.010  &  1.437$\pm$0.018 \\

CNN (Bi-Cubic) &0.689$\pm$0.012  &  1.451$\pm$0.020 \\ 

CNN+SSPP &0.692$\pm$0.017  & 1.455$\pm$0.022 \\\hline
LightViT (Crop) 	&0.635$\pm$0.015  &  1.208$\pm$0.022 \\

LightViT (Padding 0) &0.708$\pm$0.011  &  1.475$\pm$0.015 \\

LightViT (Bi-Cubic) &0.702$\pm$0.012  &  1.492$\pm$0.018 \\\hline

\textbf{Ours}       & \textbf{0.742}$\pm$0.012  &  \textbf{1.581}$\pm$0.020  \\\hline

Ours with cross embedding 	&0.720$\pm$0.013  &  1.550$\pm$0.021 \\

Ours w/o Pair-wise Disentangle 	& 0.724$\pm$0.010  &  1.548$\pm$0.019 \\
Ours w/o GAN      	&  {0.729}$\pm$0.011  &   1.546$\pm$0.020 \\\hline
\end{tabular}} 
\label{tabel:1}  
\end{table}

In our implementation, we set $P_x=1$ and $P_y=20$, i.e., $d=20$. Our encoder consisted of four PLT encoder layers with SSPP, to extract a feature $f_i$ with the size of $8\times8\times20$. Specifically, the first $8\times8\times4$ component was set as the utterance-related factors, and the remaining 16 channels were for the subject-specific factors. Then, the three 2D de-convolutional layers were applied as our decoder to generate the $64\times64$ mel-spectrogram. The activation units in our model were rectified linear units (ReLU), and we normalized the final output of each pixel using the sigmoid function. The discriminator in our model consisted of three convolutional layers and two fully connected layers, and had a sigmoid output. A detailed description of the network structure is provided in the supplementary material, due to space limitations.
  
Our model was implemented using PyTorch and trained 200 epochs for approximately 6 hours on a server equipped with an NVIDIA V100 GPU. Notably, the inference from a $\bf{H}$ matrix to audio took less than 1 second, depending on the size of $\bf{H}$. Also, the pairwise utterance consistency and GAN training were only applied during the training phase and did not affect inference. For our method and its ablation studies, we consistently set the learning rates of our heterogeneous translator and discriminator to $lr^{\mathcal{T}}=10^{-3}$ and $lr^{\mathcal{D}}=10^{-4}$, respectively, with a momentum of 0.5. The loss trade-off hyperparameters were set as $\beta=0.75$, and we set $\lambda=1$.

 It is important to note that without NMF, generating intelligible audio with a small number of subjects using video-based audio translation models, such as Lip2AudSpect \cite{akbari2018lip2audspec}, is not feasible. As an alternative, we pre-processed the input by cropping, padding with zeros, or using bi-cubic interpolation to obtain a fixed-size input $\bf{H}$. We then compared the performance of our encoder module with conventional CNN or LightViT \cite{huang2022lightvit}.

%Of note, without NMF, we can not generate intelligible audio with such few subjects with the video-based audio translation models, e.g.,~Lip2AudSpect \cite{akbari2018lip2audspec}. Instead, we can crop, padding zero, or use bi-cubic interpolation as pre-processing to get a fixed size input $\bf{H}$. Then, the conventional CNN or LightViT~\cite{huang2022lightvit} can be applied as our encoder module for comparison. 

%A possible solution to achieve the translation would be the use of convolutional neural networks (CNNs), which rely on hierarchical structures to expand their reception field and capture the complex relationships between the weighting map and audio waveform. However, this approach can result in relatively large multilayer models for our wide weighting map. Although recent window-based transformers~\cite{chu2021twins,chu2021twins,fang2022msg,yang2021focal,huang2022lightvit} have shown promise in achieving global modeling in each layer with relatively lightweight computation, their position embedding is not compatible with varied input sizes. This is particularly important for our work, as position cues are essential for motion analysis in natural sounds.

\begin{figure}[t!]
\begin{center} 
\includegraphics[width=1\linewidth]{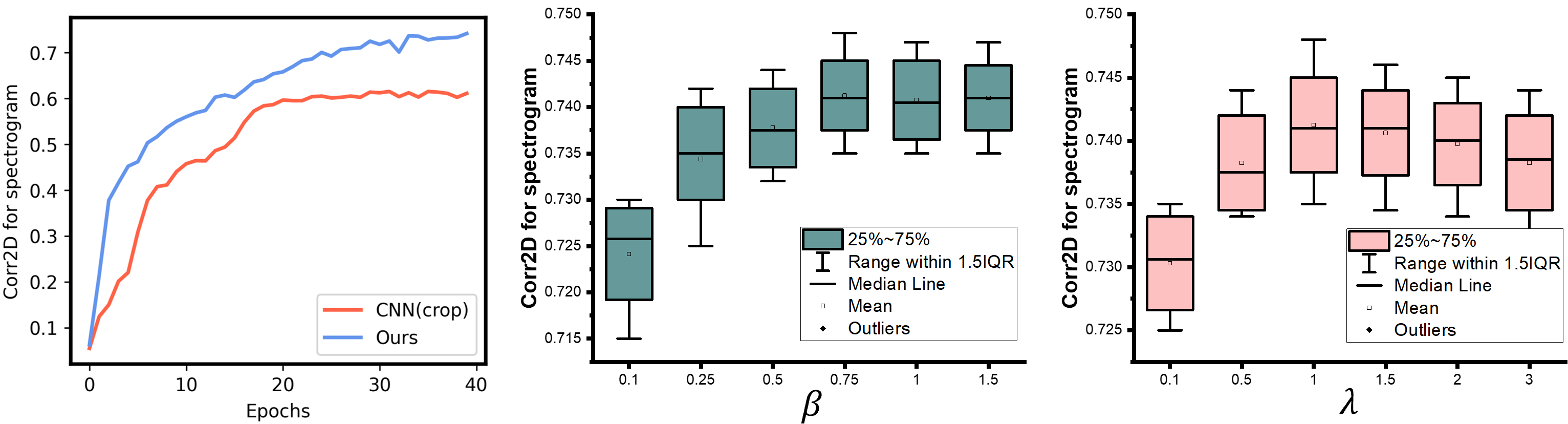}
\end{center}   
\caption{(a) Comparison of Corr2D using our plastic light transformer and CNN with crop. Sensitivity analysis of $\beta$ (b) and $\lambda$ (c).} 
\label{exp2} 
\end{figure}

Figure~\ref{fig:results} shows a qualitative comparison of our PLT framework with CNN and LightViT~\cite{huang2022lightvit} using bi-cubic interpolation. We can observe that our generated spectrogram and the corresponding audio waveforms demonstrate superior alignment with the ground truth. It is worth noting that the CNN model or the CNN-based global modeling ViTs~\cite{chu2021twins,yang2021focal} require deep models to achieve large receptive fields~\cite{luo2016understanding,araujo2019computing}. Moreover, the interpolation process adds significant computational complexity for both CNN and LightViT, making it difficult to train on a limited dataset. In Fig. \ref{exp2}(a), we show that our proposed PLT framework achieves a stable performance gain along with the training and outperforms CNN with the crop, which lost the information of some functional units.

Following \cite{akbari2018lip2audspec}, we used 2D Pearson’s correlation coefficient (Corr2D)~\cite{chi2005multiresolution}, and Perceptual Evaluation of Speech Quality (PESQ)~\cite{recommendation2001perceptual} as our evaluation metrics to measure the synthesis quality of spectrograms in the frequency domain, and waveforms in the time domain, respectively. The numerical comparisons of different encoder structures with conventional CNN or LightViT with different crop or padding strategies and our PLT framework are provided in Table \ref{tabel:1}. The standard deviation was obtained from three independent random trials. Our framework outperformed CNN and lightViT consistently. In addition, the synthesis performance was improved by pair-wise disentangled utterance consistency MMD loss and GAN loss, as demonstrated in our ablation studies. Furthermore, it outperformed the in-directional cross relative position bias~\cite{wu2021rethinking}, since two dimensions in the weighting map indicate time and spatial relationship, respectively. Notably, even though CNN with SSPP can process varied size inputs, it suffers from limited long-term modeling capacity~\cite{luo2016understanding,araujo2019computing} and unstable performance~\cite{richter2021input}. The sensitivity analysis of our loss weights are given in Fig. \ref{exp2}(b) and (c), where the performance was relatively stable for $\beta\in[0.75,1.5]$ and $\lambda\in[1,2]$.

\section{Conclusion}
 
This work aimed to explore the relationship between tongue movements and speech acoustics by translating weighting maps, which represent the functional units of the tongue, to their corresponding audio waveforms. To achieve this, we proposed a deep PLT framework that can handle variable-sized weighting maps and generated fixed-sized spectrograms, without information loss or dimension expansion. Our framework efficiently modeled global correlations in wide matrix input. To improve the realism of the generated spectrograms, we applied pair-wise utterance consistency with MMD constraint and adversarial training. Our experimental results demonstrated  the potential of our framework to synthesize audio waveforms from weighting maps, which can aid clinicians and researchers in better understanding the relationship between the two modalities.

\section*{Acknowledgements}
 
This work is supported by NIH R01DC014717, R01DC018511, R01CA133015, and P41EB022544.

%In this work, we proposed a novel framework to synthesize spectrograms from tagged-MRI sequences. The audio waveforms can also be obtained from the synthesized spectrograms. In particular, we proposed an efficient fully convolutional asymmetry translator with help of a self residual attention strategy to specifically focus on the moving muscular structures for speech production.~Additionally, we used a pairwise correlation of the samples with the same utterances with a latent space representation disentanglement scheme.~Furthermore, we incorporated an adversarial training approach with GAN to yield improved results on our generated spectrograms.~Our experimental results showed that our framework was able to successfully synthesize spectrograms (and clear waveforms) from tagged-MRI sequences, outperforming the Lipreading based method. Therefore, our framework offered the potential to help clinicians improve treatment strategies for patients with speech-related disorders. In future work, we will investigate the use of full three-dimensional plus time tagged-MRI sequences as well as tracking information from tagged-MRI to achieve spectrogram synthesis.

%\section*{Acknowledgements}

%This work is partially supported by NIH R01DC018511, R01DE027989, and P41EB022544.

%\section*{Acknowledgements}
 
%This work is supported by NIH R01DC014717, R01DC018511, and R01CA133015.

\bibliographystyle{splncs04}
\bibliography{egbib}

\begin{thebibliography}{10}
\providecommand{\url}[1]{\texttt{#1}}
\providecommand{\urlprefix}{URL }
\providecommand{\doi}[1]{https://doi.org/#1}

\bibitem{akbari2018lip2audspec}
Akbari, H., Arora, H., Cao, L., Mesgarani, N.: Lip2{audspec}: Speech
  reconstruction from silent lip movements video. In: ICASSP. pp. 2516--2520.
  IEEE (2018)

\bibitem{araujo2019computing}
Araujo, A., Norris, W., Sim, J.: Computing receptive fields of convolutional
  neural networks. Distill  \textbf{4}(11), ~e21 (2019)

\bibitem{che2019deep}
Che, T., Liu, X., Li, S., Ge, Y., Zhang, R., Xiong, C., Bengio, Y.: Deep
  verifier networks: Verification of deep discriminative models with deep
  generative models. AAAI  (2021)

\bibitem{chi2005multiresolution}
Chi, T., Ru, P., Shamma, S.A.: Multiresolution spectrotemporal analysis of
  complex sounds. The Journal of the Acoustical Society of America
  \textbf{118}(2),  887--906 (2005)

\bibitem{chu2021twins}
Chu, X., Tian, Z., Wang, Y., Zhang, B., Ren, H., Wei, X., Xia, H., Shen, C.:
  Twins: Revisiting the design of spatial attention in vision transformers.
  Advances in Neural Information Processing Systems  \textbf{34},  9355--9366
  (2021)

\bibitem{chung2016lip}
Chung, J.S., Zisserman, A.: Lip reading in the wild. In: ACCV. pp. 87--103.
  Springer (2016)

\bibitem{dosovitskiy2020image}
Dosovitskiy, A., Beyer, L., Kolesnikov, A., Weissenborn, D., Zhai, X.,
  Unterthiner, T., Dehghani, M., Minderer, M., Heigold, G., Gelly, S., et~al.:
  An image is worth 16x16 words: Transformers for image recognition at scale.
  arXiv preprint arXiv:2010.11929  (2020)

\bibitem{dziugaite2015training}
Dziugaite, G.K., Roy, D.M., Ghahramani, Z.: Training generative neural networks
  via maximum mean discrepancy optimization. arXiv preprint arXiv:1505.03906
  (2015)

\bibitem{ephrat2017vid2speech}
Ephrat, A., Peleg, S.: Vid2speech: speech reconstruction from silent video. In:
  ICASSP. pp. 5095--5099. IEEE (2017)

\bibitem{goodfellow2020generative}
Goodfellow, I., Pouget-Abadie, J., Mirza, M., Xu, B., Warde-Farley, D., Ozair,
  S., Courville, A., Bengio, Y.: Generative adversarial networks.
  Communications of the ACM  \textbf{63}(11),  139--144 (2020)

\bibitem{griffin1984signal}
Griffin, D., Lim, J.: Signal estimation from modified short-time fourier
  transform. IEEE Transactions on acoustics, speech, and signal processing
  \textbf{32}(2),  236--243 (1984)

\bibitem{he2020image2audio}
He, G., Liu, X., Fan, F., You, J.: Image2audio: Facilitating semi-supervised
  audio emotion recognition with facial expression image. In: Proceedings of
  the IEEE/CVF Conference on Computer Vision and Pattern Recognition Workshops.
  pp. 912--913 (2020)

\bibitem{he2015spatial}
He, K., Zhang, X., Ren, S., Sun, J.: Spatial pyramid pooling in deep
  convolutional networks for visual recognition. IEEE transactions on pattern
  analysis and machine intelligence  \textbf{37}(9),  1904--1916 (2015)

\bibitem{huang2022lightvit}
Huang, T., Huang, L., You, S., Wang, F., Qian, C., Xu, C.: Lightvit: Towards
  light-weight convolution-free vision transformers. arXiv preprint
  arXiv:2207.05557  (2022)

\bibitem{lee2013semi}
Lee, J., Woo, J., Xing, F., Murano, E.Z., Stone, M., Prince, J.L.:
  Semi-automatic segmentation of the tongue for 3{D} motion analysis with
  dynamic {MRI}. In: ISBI. pp. 1465--1468. IEEE (2013)

\bibitem{liu2021mutual}
Liu, X., Chao, Y., You, J.J., Kuo, C.C.J., Vijayakumar, B.: Mutual information
  regularized feature-level frankenstein for discriminative recognition. IEEE
  TPAMI  (2021)

\bibitem{liu2021domain}
Liu, X., Hu, B., Jin, L., Han, X., Xing, F., Ouyang, J., Lu, J., Fakhri, G.E.,
  Woo, J.: Domain generalization under conditional and label shifts via
  variational {B}ayesian inference. IJCAI  (2021)

\bibitem{liu2019feature}
Liu, X., Li, S., Kong, L., Xie, W., Jia, P., You, J., Kumar, B.: Feature-level
  frankenstein: Eliminating variations for discriminative recognition. In:
  CVPR. pp. 637--646 (2019)

\bibitem{liu2017adaptive}
Liu, X., Vijaya~Kumar, B., You, J., Jia, P.: Adaptive deep metric learning for
  identity-aware facial expression recognition. In: CVPR. pp. 20--29 (2017)

\bibitem{liu2021swin}
Liu, Z., Lin, Y., Cao, Y., Hu, H., Wei, Y., Zhang, Z., Lin, S., Guo, B.: Swin
  transformer: Hierarchical vision transformer using shifted windows. In:
  Proceedings of the IEEE/CVF international conference on computer vision. pp.
  10012--10022 (2021)

\bibitem{luo2016understanding}
Luo, W., Li, Y., Urtasun, R., Zemel, R.: Understanding the effective receptive
  field in deep convolutional neural networks. Advances in neural information
  processing systems  \textbf{29} (2016)

\bibitem{recommendation2001perceptual}
Recommendation, I.T.: Perceptual evaluation of speech quality {PESQ}): An
  objective method for end-to-end speech quality assessment of narrow-band
  telephone networks and speech codecs. Rec. ITU-T P. 862  (2001)

\bibitem{richter2021input}
Richter, M.L., Byttner, W., Krumnack, U., Wiedenroth, A., Schallner, L., Shenk,
  J.: (input) size matters for cnn classifiers. In: 30th International
  Conference on Artificial Neural Networks. pp. 133--144. Springer (2021)

\bibitem{salimans2016improved}
Salimans, T., Goodfellow, I., Zaremba, W., Cheung, V., Radford, A., Chen, X.:
  Improved techniques for training gans. NIPS  \textbf{29},  2234--2242 (2016)

\bibitem{woo2018sparse}
Woo, J., Prince, J.L., Stone, M., Xing, F., Gomez, A.D., Green, J.R., Hartnick,
  C.J., Brady, T.J., Reese, T.G., Wedeen, V.J., et~al.: A sparse non-negative
  matrix factorization framework for identifying functional units of tongue
  behavior from mri. IEEE transactions on medical imaging  \textbf{38}(3),
  730--740 (2018)

\bibitem{woo2021deep}
Woo, J., Xing, F., Prince, J.L., Stone, M., Gomez, A.D., Reese, T.G., Wedeen,
  V.J., El~Fakhri, G.: A deep joint sparse non-negative matrix factorization
  framework for identifying the common and subject-specific functional units of
  tongue motion during speech. Medical image analysis  \textbf{72},  102131
  (2021)

\bibitem{woo2020identifying}
Woo, J., Xing, F., Prince, J.L., Stone, M., Reese, T.G., Wedeen, V.J.,
  El~Fakhri, G.: Identifying the common and subject-specific functional units
  of speech movements via a joint sparse non-negative matrix factorization
  framework. In: Medical Imaging 2020: Image Processing. vol. 11313, pp.
  446--451. SPIE (2020)

\bibitem{wu2021rethinking}
Wu, K., Peng, H., Chen, M., Fu, J., Chao, H.: Rethinking and improving relative
  position encoding for vision transformer. In: Proceedings of the IEEE/CVF
  International Conference on Computer Vision. pp. 10033--10041 (2021)

\bibitem{xing2017phase}
Xing, F., Woo, J., Gomez, A.D., Pham, D.L., Bayly, P.V., Stone, M., Prince,
  J.L.: Phase vector incompressible registration algorithm for motion
  estimation from tagged magnetic resonance images. IEEE TMI  \textbf{36}(10)
  (2017)

\bibitem{xing20133d}
Xing, F., Woo, J., Murano, E.Z., Lee, J., Stone, M., Prince, J.L.: {3D} tongue
  motion from tagged and cine {MR} images. In: MICCAI. pp. 41--48. Springer
  (2013)

\bibitem{yang2021focal}
Yang, J., Li, C., Zhang, P., Dai, X., Xiao, B., Yuan, L., Gao, J.: Focal
  self-attention for local-global interactions in vision transformers. arXiv
  preprint arXiv:2107.00641  (2021)

\bibitem{zhang2021rest}
Zhang, Q., Yang, Y.B.: Rest: An efficient transformer for visual recognition.
  Advances in Neural Information Processing Systems  \textbf{34},  15475--15485
  (2021)

\end{thebibliography}

\end{document}